\documentclass[11pt,a4paper]{article}
\usepackage[left=2.5cm,top=2.5cm,right=2.5cm,bottom=2.5cm]{geometry}
\usepackage{graphicx}
\usepackage{bm}
\usepackage{doi}
\usepackage{color}

\usepackage{comment}

\begin{document}

\clearpage
\noindent \textbf{\LARGE Doping of metal--organic frameworks towards resistive sensing}\\
\vspace{0cm}
\noindent \textbf{Hidetsugu Shiozawa,$^{1\ast}$ Bernhard C. Bayer,$^{1}$ Herwig Peterlik,$^{1}$ Jannik C. Meyer,$^{1}$ Wolfgang Lang,$^{1}$ Thomas Pichler$^{1}$}\\
\noindent $^{1}$Faculty of Physics, University of Vienna, Boltzmanngasse 5, 1090 Vienna, Austria\\
\noindent $^\ast$To whom correspondence should be addressed; E-mail: hidetsugu.shiozawa@univie.ac.at\\
\hspace{5mm}
\begin{abstract}
Coordination polymerization leads to various metal--organic frameworks (MOFs) in which symmetrical metal nodes exposed to nano voids lead to unique physical properties and chemical functionalities. One of the challenges towards their applications as porous materials is to make MOFs optimally conductive to be used as electronic components. Here, we demonstrate that Co-MOF-74, a honeycomb nano--framework with one--dimensionally arranged cobalt atoms advances its physical properties by accommodating Tetracyanochinodimethan (TCNQ), an acceptor molecule. Strong intermolecular charge transfer narrows the optical band gap down to 1.5~eV of divalent TCNQ and enhances the electrical conduction, which allows the MOF to be utilized for resistive gas- and photo-sensing. Our result provides insight into electronic interactions in doped MOFs and paves the way towards their electronic applications.
\end{abstract}

\section*{Introduction}
Metal--organic frameworks (MOFs) \cite{Yaghi03N,Kitagawa04ACE} are exciting materials due to their unprecedented nanostructures that are tailored on a bulk scale by matching up designated metal ligands and organic linkers.
Their colourful appearances that can change for instance with solvent exchange and molecular doping \cite{Allendorf15JoPCL}
signify their chromophoric nature.
Also, MOFs' diverse physical and chemical tunability is promising as demonstrated via building block replacement \cite{Deria14CSR} and infiltration with redox-active molecules \cite{Allendorf15JoPCL}.
MOFs' high internal surface areas are advantageous for gas sorption \cite{Li12CR}
or chemical sensors \cite{Kreno12CR},
as demonstrated with H$_2$ \cite{Rosi03S},
CO$_2$ \cite{Sumida12CR},
N$_2$ and methane \cite{Furukawa10S}.
MOFs built with TCNQ linker exhibited an excellent selective uptake of O$_2$ and NO \cite{Shimomura10NC} in comparison to N$_2$, CO, CO$_2$, C$_2$H$_2$ or Ar, and separation of benzene from cyclohexane \cite{Shimomura07JotACS}, as a result of guest-host charge transfer interactions.
Furthermore, nanoscale voids in MOFs were able to accommodate molecules as large as bucky fullerene
\cite{Chae04N,Inokuma10NC}.

One of the challenges towards their electronic applications is to make MOFs optimally conductive to be used as electronic components \cite{Stavila14CSR}. It is only recently that electrically conductive MOFs were realized through covalent interactions \cite{Sheberla14JotACS}
or redox active guest molecules, i.e., hole doping with TCNQ\cite{Talin14S}.

In this article, we study Co-MOF-74 (also called Co-CPO-27 or Co2(dobdc)) crystals \cite{Rosi05JotACS} in various forms from nanocrystals to microcrystals and thin films (see SI, section 1.1). Infiltrated with TCNQ (see the diagram in Fig.\ref{XRD}a), charge transfer between the host framework and the guest molecule leads to advanced optical and electrical transport properties. Based on the enhanced electronic conduction of the charge-transfer MOF as well as its extended optical absorption profile covering the entire visible range, we demonstrate their resistive photo- and gas-sensing capabilities, both of which are key properties towards MOFs' electronic applications.

\vspace{12pt}
\section*{Results and discussion}

\subsection*{Structure}
\begin{figure}[h]
\centering
\includegraphics[width=140mm]{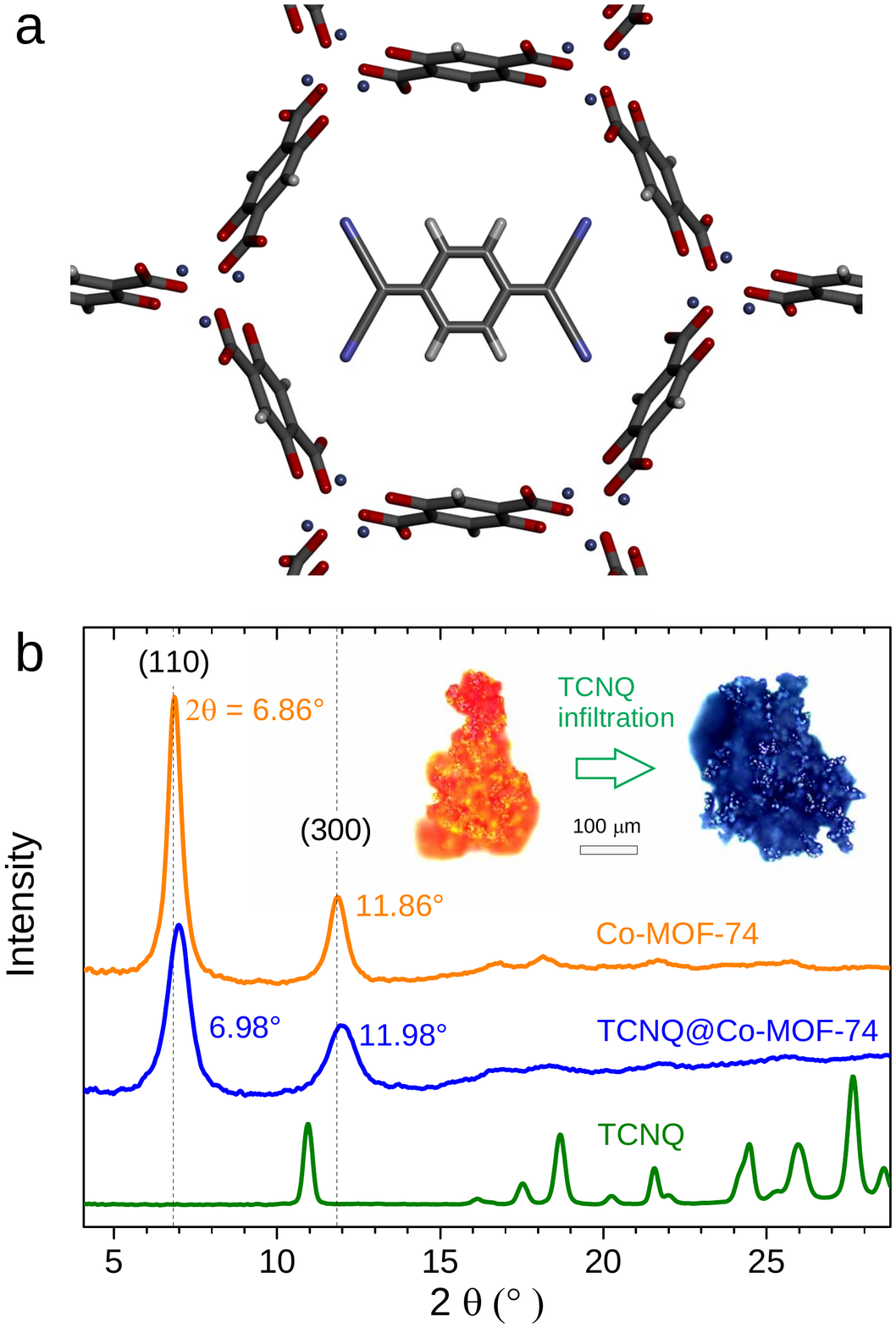}
\caption{
{\bf a}.A diagram of a Co-MOF-74 framework encapsulating a TCNQ molecule.
{\bf b}.X-ray diffraction profiles for Co-MOF-74 and Co-MOF-74 microcrystallites infiltrated with TCNQ in comparison to TCNQ powder. The dashed vertical lines are the first two diffraction lines expected for Co-MOF-74 \cite{Dietzel05ACE}. Inset: optical micrographs of microcrystallites before (left) and after (right) the infiltration with TCNQ.
}
\label{XRD}
\end{figure}

X-ray diffraction profiles for Co-MOF-74 microcrystals before and after infiltration with TCNQ are plotted with respect to $2 \Theta$ in Fig.~\ref{XRD}b. The peaks at 6.86$^{\circ}$ and 11.86$^{\circ}$ indexed to (1~1~0) and (3~0~0) with $\Theta_{(3~0~0))}/\Theta_{(1~1~0)} \sim \sqrt{3}$ originate from the hexagonal structure. The dashed vertical lines are the first two diffraction lines expected for Co-MOF-74 \cite{Dietzel05ACE}. The corresponding lattice spacings are 0.645 and 0.375~nm, respectively. With the inclusion of TCNQ, the peaks get slightly upshifted and broader, indicating a contracted and distorted lattice structure, respectively (see SI, section 3).

\begin{figure}[h]
\centering
\includegraphics[width=150mm]{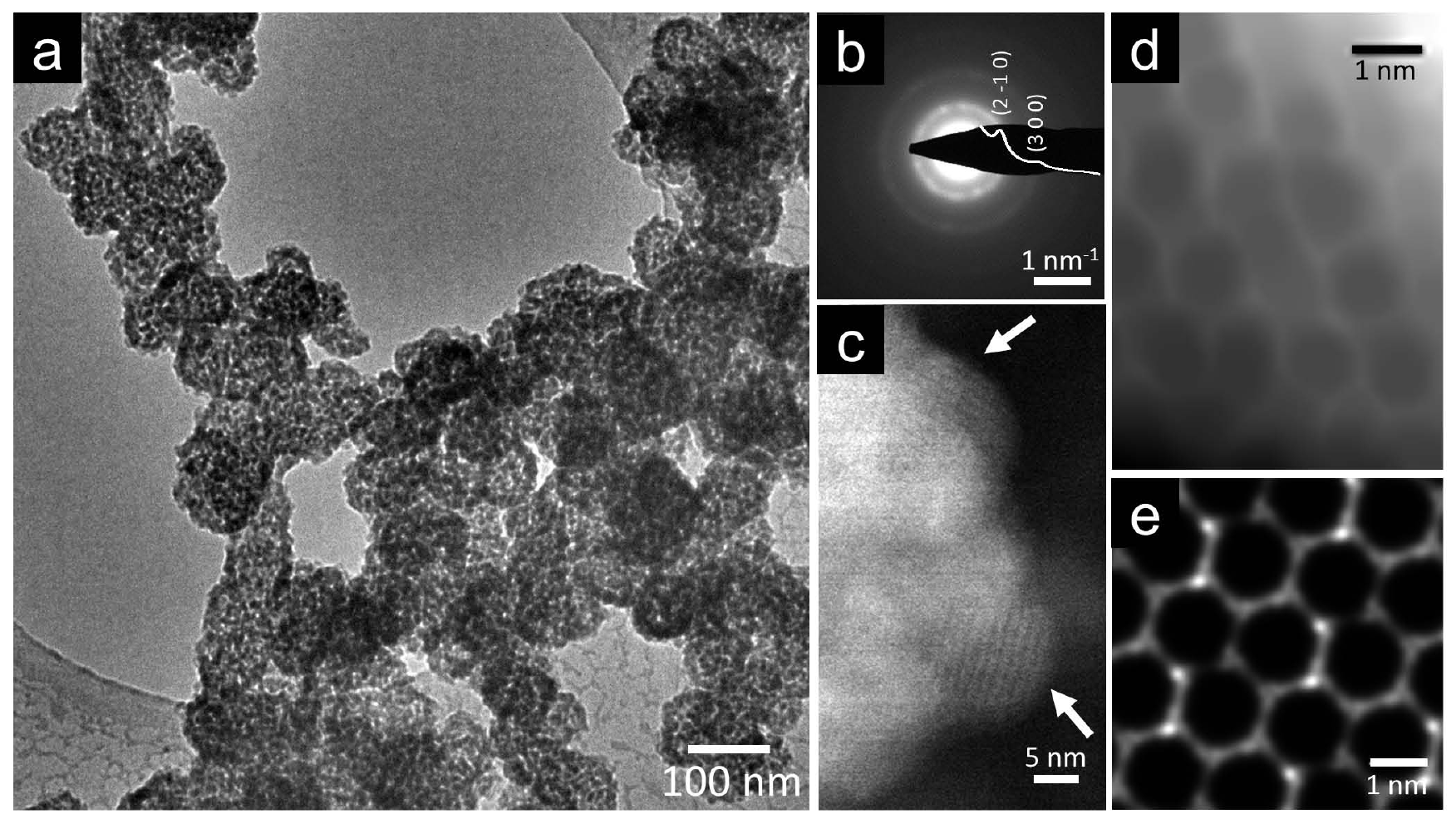}
\caption{
{\bf a}. BF TEM image of agglomerated Co-MOF-74 nanocrystals. {\bf b}. SAED pattern of such nanocrystals in which the extracted radial intensity profile in the overlay confirms their Co-MOF-74 structure \cite{Dietzel05ACE}. {\bf c}. ADF STEM image of a nanocrystal agglomerate such as in {\bf a} at higher magnification. The observed lattice fringes within the individual nanocrystals (marked by arrows) are consistent with Co-MOF-74. {\bf d}. ADF STEM image (Gaussian blurred and minimum filtered) consistent with the pore structure of a Co-MOF-74 nanocrystal as in {\bf c}. {\bf e}. STEM image simulation of a Co-MOF-74 crystal viewed along the [001] zone axis (rotated by -18$^\circ$ in plane to match {\bf d}). Further details on TEM/STEM results are given in SI, section 2.
}
\label{tem}
\end{figure}

A bright field (BF) transmission electron microscopy (TEM) image of Co-MOF-74 in Fig. \ref{tem}a shows the assemblage of particles (50--180~nm).
The selected area electron diffraction (SAED) in Fig.~\ref{tem}b confirms the crystallite Co-MOF-74 structure, consistent with the XRD profile in Fig.~\ref{XRD}b. Annular dark field (ADF) scanning transmission electron microscopy (STEM) at higher magnification in Fig.~\ref{tem}c further corroborates the crystalline nature of individual nanocrystals of $\sim$4--15~nm size.
Figure~\ref{tem}d finally visualizes the honeycomb pore structure in one of these Co-MOF-74 nanocrystals, consistent with a Co-MOF-74 crystal viewed along the [001] zone axis, as shown in the image simulation in Fig.~\ref{tem}e.
Note that characterization of MOFs by TEM/STEM is notoriously challenging because of the high susceptibility of the organic linkers to electron beam induced degradation \cite{Diaz-Garcia14CG&D,Mayoral15C}. Our high-resolution microscopy data is the first such available on Co-MOF-74 nanocrystals to date \cite{Diaz-Garcia14CG&D,Mayoral15C,Gimeno-Fabra12CC}.

\subsection*{Optical absorption spectroscopy}
\begin{figure}[h]
\centering
\includegraphics[width=200mm]{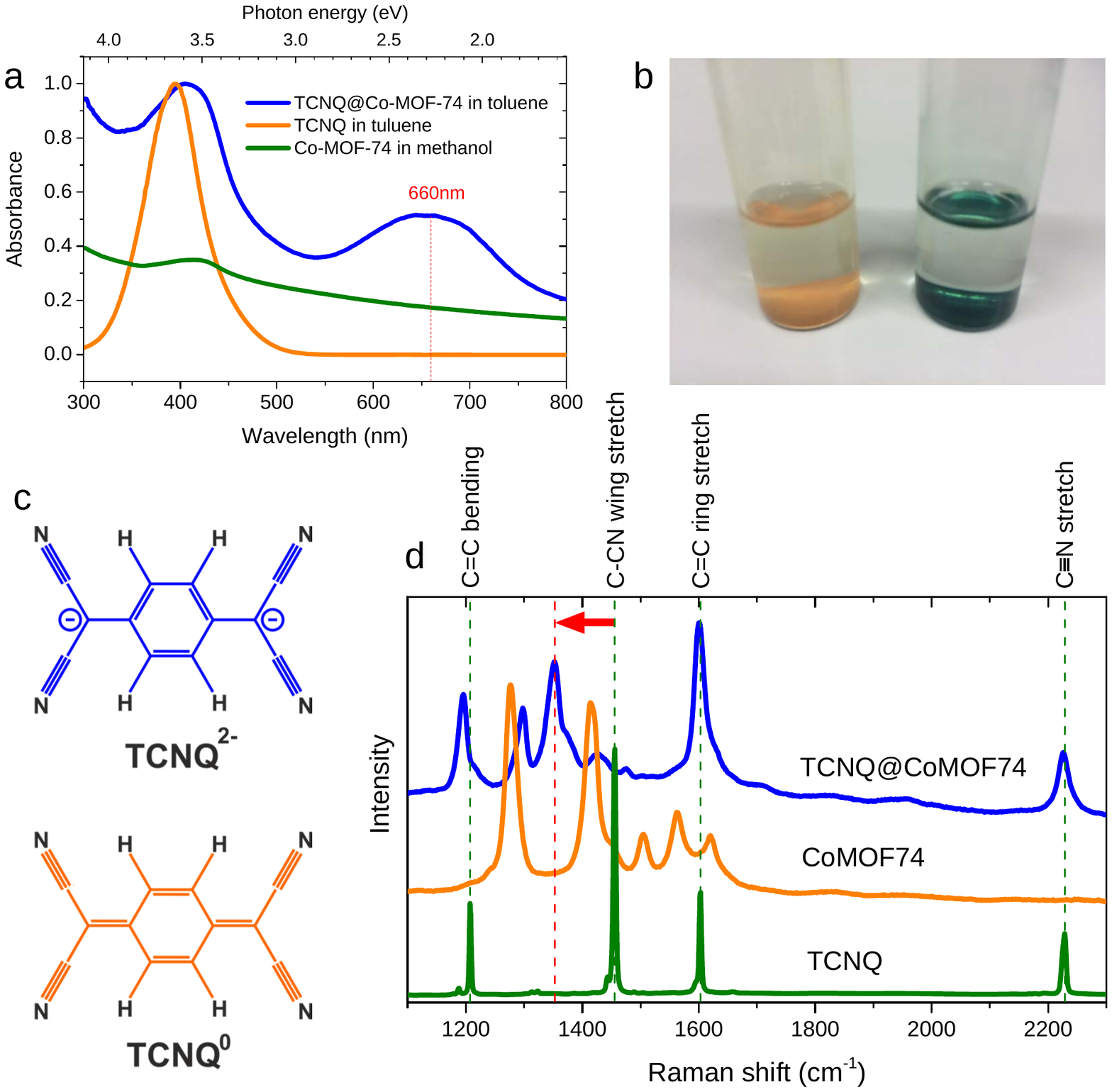}
\caption{{\bf a}. Absorption spectra for TCNQ, Co-MOF-74 and TCNQ@Co-MOF-74 nanocrystals.
{\bf b}. A photograph of the glass vials containing Co-MOF-74 (left) and TCNQ@Co-MOF-74 (right) nanocrystals, both in methanol.
{\bf c}. Diagrams for neutral TCNQ$^0$ (bottom) and divalent TCNQ$^{2-}$ (top).
{\bf d}. Raman spectra for TCNQ, Co-MOF-74 and Co-MOF-74 microcrystals infiltrated with TCNQ collected at 633~nm laser wavelength.
The dashed vertical lines in green are the frequencies for the four major Raman lines of TCNQ$^0$, while that in red is associated to the frequency for red-shifted C-CN wing stretch mode of TCNQ$^{2-}$ in the Co-MOF-74.}
\label{OptRaman}
\end{figure}

A prominent change upon infiltration with TCNQ is observed in the MOF's colour that is originally orange turning into dark cyan, see the optical micrographs in Fig.~\ref{OptRaman}. As shown in Fig. \ref{OptRaman}a, neutral TCNQ$^0$ in toluene has its giant optical transition peak located at about $\sim$395~nm (3.15~eV), that makes it look orange, while Co-MOF-74 has a weak absorption peak at $\sim$410~nm (3.0~eV). As TCNQ infiltrates Co-MOF-74, a strong new absorption peak emerges at $\sim$660~nm (1.9~eV) in addition to the Co-MOF-74 peak. Accordingly, the optical gap estimated from the lowest energy peak onset is reduced by more than 1.0~eV. The strongest absorption structures of TCNQ$^{-}$ monomer at around 800~nm \cite{BOYD65JoCP,HIROMA71BotCSoJ,Oohashi73BotCSoJ,TORRANCE79PRB} are not visible, excluding the presence of monovalent TCNQ.
Absorption at $\sim$ 2~eV ($\sim$640~nm) was previously observed as a minor component of absorption profiles for TCNQ$^{-}$ dimers in aqueous solution \cite{BOYD65JoCP,HIROMA71BotCSoJ,Oohashi73BotCSoJ,TORRANCE79PRB}, blue-shifted as compared to the corresponding absorption in TCNQ$^{-}$ monomers. Absorption structures at a similar energy region were also observed in TCNQ$^{-}$ salts in the solid state \cite{TORRANCE79PRB},
but again only as a minor absorption associated with locally excited transitions polarized perpendicular to the molecular chain axis \cite{HIROMA71BotCSoJ}.

It is known that the disproportionation occurs so a TCNQ$^-$ dimer transforms to a pair of TCNQ$^0$ and TCNQ$^{2-}$ \cite{Hashimoto16CPL}.
The TCNQ$^{2-}$ divalent anion in solutions exhibits a large broad absorption peak at around 2~eV ($\sim$640~nm) \cite{Hashimoto16CPL},
along with high energy absorption peaks at about 300~nm \cite{KHATKALE79JoCP,SUCHANSKI76JotACS}.
Hence, the 2~eV ($\sim$640~nm) peak of TCNQ$^-$ dimers observed previously could be the component of disproportionated TCNQ$^{2-}$.
Note that TCNQ$^{2-}$'s oxidation product, dicyano-p-toluoyl cyanide (DCTC$^{-}$), has an absorption peak at about 480~nm as observed via electrochemical doping \cite{JONKMAN72CPL,SUCHANSKI76JotACS} and molecular orbital calculations \cite{BIEBER74CP}, as well as in charge transfer salts \cite{Grossel00CoM}.

In our case, the 660~nm peak can be attributed to that of TCNQ$^{2-}$, which means that the charge transfer occurs so [TCNQ]$^{2-}$ [Co-MOF-74]$^{2+}$ salt is created.

\subsection*{Raman spectroscopy}
The assignment of the 660~nm peak to the TCNQ is backed up by the fact that Raman response of the TCNQ acceptor is resonance enhanced, see SI, section 5.
Fig.\ref{OptRaman}b compares Raman data collected with a laser wavelength of 633~nm (in resonance with TCNQ$^{2-}$) for Co-MOF-74 before and after infiltration with TCNQ as well as that for TCNQ crystals.
The major four lines for TCNQ crystals are located at Raman frequencies of 1207.5, 1455.5, 1602.8 and 2228.5~cm$^{-1}$, assigned to the C=CH bending, C-CN wing stretch, C=C ring stretch and C$\equiv$N stretch modes, respectively. In TCNQ@Co-MOF-74, those except for the C-CN wing stretch mode are visible at 1195.5, 1600.8, 2226.0~cm$^{-1}$ although the C=CH bending mode frequency is markedly red-shifted by -12~cm$^{-1}$. There are weak peaks that can be assigned as frequency-shifted modes for the Co-MOF-74 component, except for the peak at 1352~cm$^{-1}$ whose intensity is comparable to those of the other three TCNQ modes. Hence, this peak can be assigned to the C-CN wing stretch mode shifted by -103.5~cm$^{-1}$. Indeed, the C-CN wing stretch mode is known to frequency-shift by doping as electrons are preferentially accommodated in the wings of TCNQ \cite{MILLER87JoPC}. From a comparison to previously reported frequency shifts for ionic TCNQs,
the corresponding doping level is $\sim$ 1.5~$\pm$~0.2~e$^-$ per TCNQ, see SI, section 6.
This non-integer charge number lower than 2~e$^-$ per TCNQ estimated from the optical absorption data in the previous section could be due to possible guest-host orbital mixing, as reported previously in Cu$_3$(BTC)$_2$ (BTC = benzene tricarboxylate), commonly known as HKUST-1 \cite{Talin14S}.
Note that no sign of DCTC$^{-}$ is observed in the Raman data \cite{SUCHANSKI76JotACS,KHATKALE79JoCP,HARRIS95VS}.

\subsection*{XPS}
\begin{figure}[h]
\centering
\includegraphics[width=120mm]{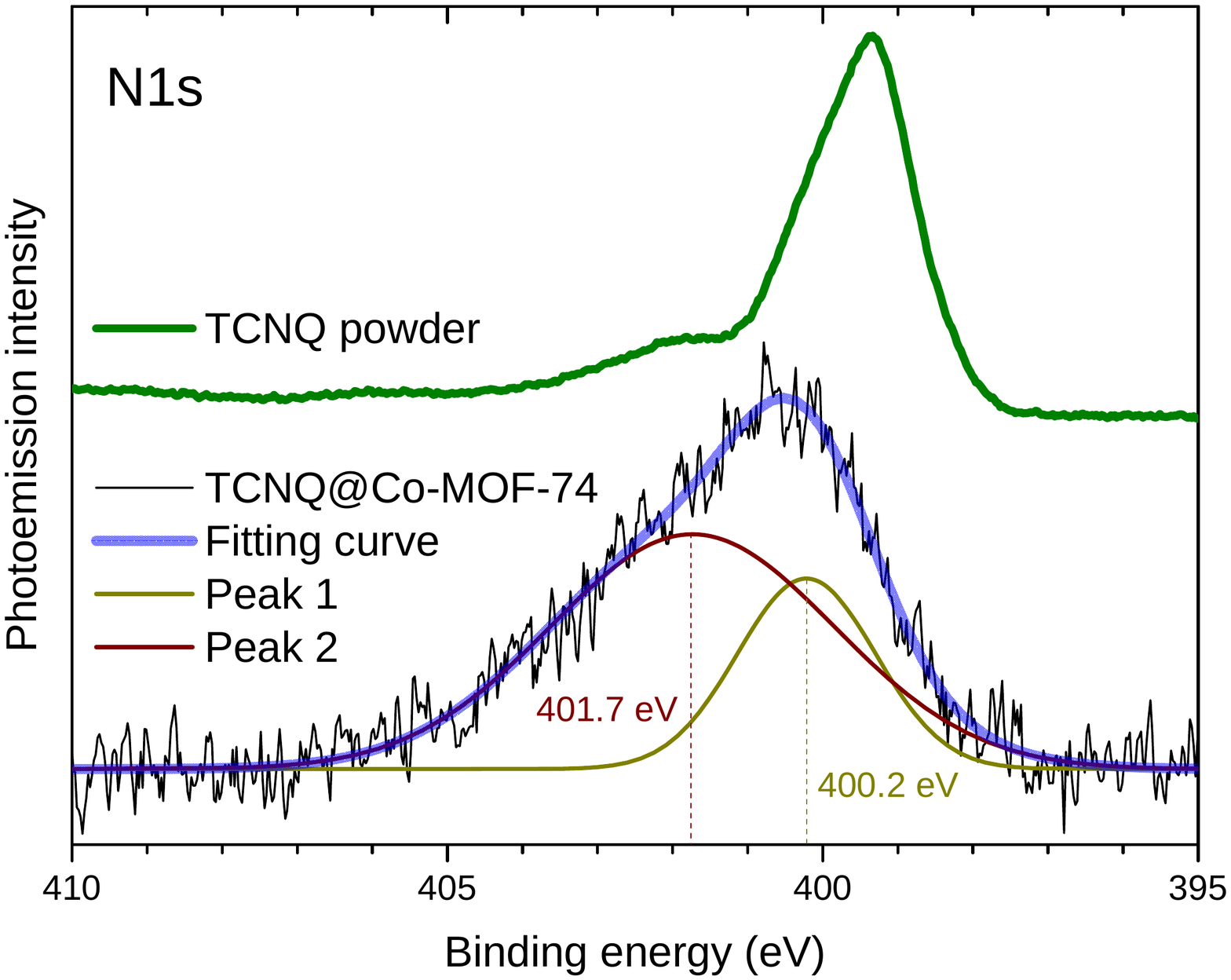}
\caption{
N1$s$ photoemission spectra for TCNQ and TCNQ@Co-MOF-74. The latter is fit with two gaussian peaks.}
\label{XPS}
\end{figure}

If there is a guest-host chemical bonding, it could be between oxygen and TCNQ wing as suggested in the copper-based HKUST-1 infiltrating TCNQ \cite{Talin14S}. It could also be such a process like TiO$_2$-TCNQ colouration leading to interfacial charge-transfer optical transitions \cite{Fujisawa15PCCP}.
Figure \ref{XPS}b shows X-ray photoemission spectroscopy (XPS) data at the N1$s$ edge. For neutral TCNQ in powder, N$1s$ peak is observed at 399.4~eV, about ~2~eV above which is a shake-up.
For the TCNQ in MOFs, the main peak (peak 1) is shifted to 400.2~eV with an enhanced broad shake-up peak (peak 2) at 401.7~eV. This is in contrast to the case for TCNQ$^{-}$/Cu(100) \cite{Tseng10NC} and lithium-intercalated TCNQ \cite{Precht15PCCP}, in which the main peak is shifted towards a lower binding energy and the shape-up feature suppressed, both attributed to the addition of one electron. Screening effect enhanced due to reduced charge transfer energy $\Delta_{CT}$ should also reduce the binding energy, that could be due to enhanced conductance in the TCNQ$^{2-}$. Hence, molecular orbital changes that possibly involve orbital hybridizations between nitrogen and the host is likely to explain the energy shifts to higher binding energies. The two N~1$s$ splits with area intensities of 0.42 (peak 1) and 0.58 (peak 2) are indicative of two distinguished chemical states of nitrogen due to an asymmetric charge distribution as a result of chemical bonding possibly on one side of the TCNQ. See SI, section 7 for further details.

\subsection*{Transport properties}
\begin{figure}[h]
\centering
\includegraphics[width=160mm]{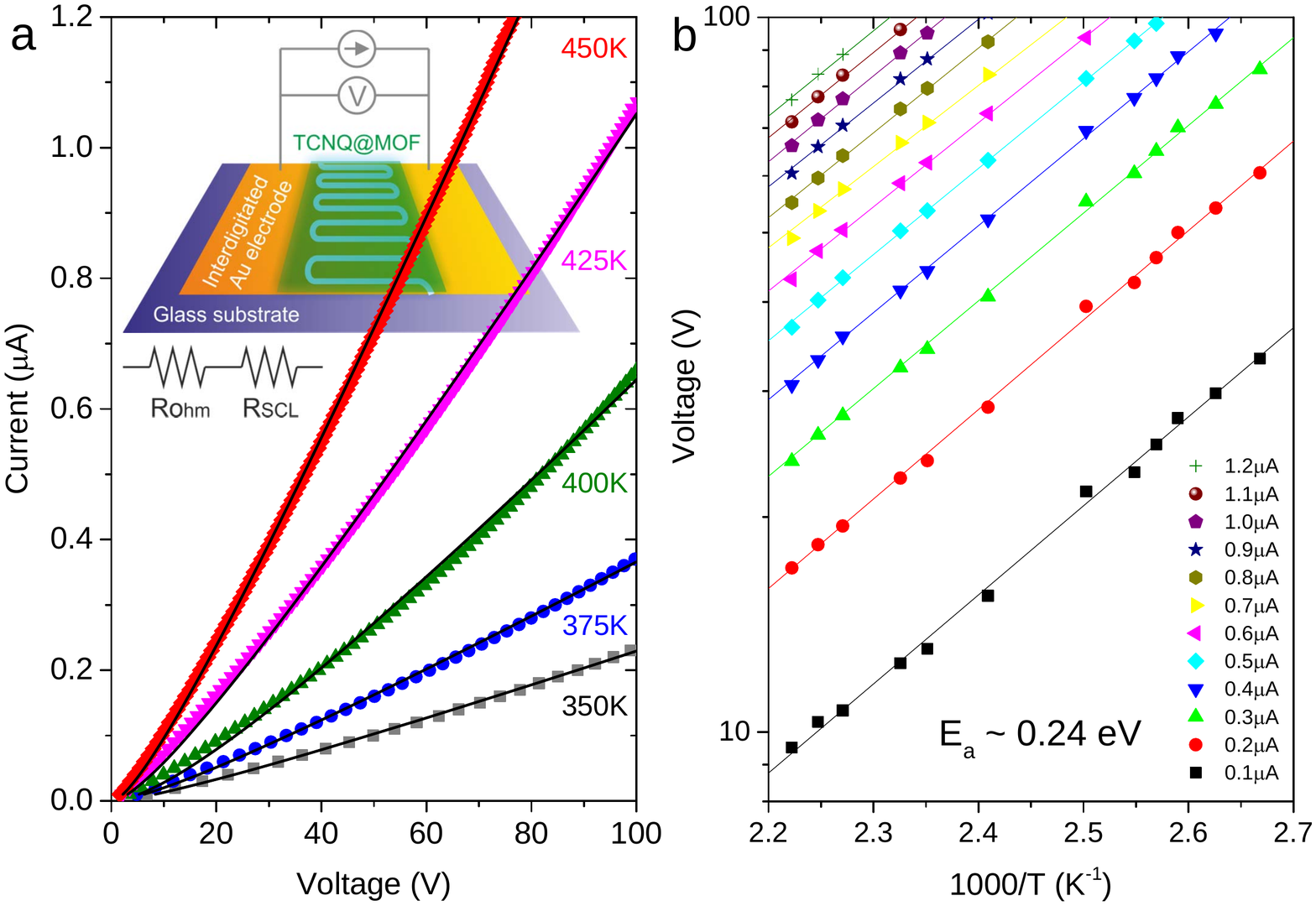}
\caption{{\bf a}. Current--voltage (I-V) data for TCNQ@Co-MOF-74 measured at 350, 375, 400, 425 and 450~K, fit to $V = aI + bI^{0.5}$ (the solid curves) assuming that the measured resistance is expressed as ohmic ($I \propto V_{Ohm}$) and space charge limited ($I \propto V_{SCL}^2$) resistances connected in series.
Inset: a diagram for transport measurements on a TCNQ@Co-MOF-74 film deposited on an interdigitated gold electrode.
{\bf b}. Voltages versus inverse temperature measured at different currents from $I$ = 0.1~$\mu$A up to 1.2~$\mu$A. The solid lines are Arrhenius functions $V^I(T) \propto$ exp$(-E_a/k_BT)$ fitting the data well with an activation energy of $E_a \sim$ 0.24~eV.}
\label{I-V}
\end{figure}

Provided that the electronic gap is reduced by doping, our charge-transfer TCNQ@Co-MOF-74 could become more conductive, as reported previously on HKUST-1 infiltrated with TCNQ \cite{Talin14S}. While the non-doped MOF is as insulating as the glass substrate at temperatures up to 450~K, the resistance of the doped MOF falls into a measurable range. Fig.~\ref{I-V}a presents the current--voltage (I-V) characteristics of a TCNQ-MOF film grown directly on interdigitated gold electrodes on a glass plate. The data measured at various temperatures in a range from 350 up to 450~K are nonlinear and follow a power-law dependence $I \propto V^{1.2}$ (see SI, section 8.1). For the electrical conduction through metal electrode--dielectric interfaces, the space charge limited (SCL) conduction model \cite{ROSE55PR} predicts a power-law scaling $I \propto V^2$. Assuming that the measured conduction across the TCNQ-MOF device is expressed as ohmic ($V_{Ohm} = aI$) and SCL ($V_{SCL} = bI^{0.5}$) resistances connected in series, i.e., $V = aI + bI^{0.5}$, we can fit the data fairly well, see Fig.~\ref{I-V}a. The ratio of $V_{Ohm}$ to $V_{SCL}$, apparently scaled by $I^{0.5}$, goes down with the temperature, but stays well above one in the sub-$\mu$A range (see SI, section 8.2), meaning that the ohmic conduction is dominant.

In turn, the voltage versus inverse temperature measured with various currents up to 1.2~$\mu$A exhibits a linear slope in an Arrhenius plot in accordance with the function $V(T) \propto exp(-E_a/k_BT)$, where $k_B$ is the Boltzmann constant, with an carrier injection activation energy of $E_a$ = 0.25~eV, see Fig.~\ref{I-V}b.

\subsection*{Sensing}

\begin{figure}[h]
\centering
\includegraphics[width=80mm]{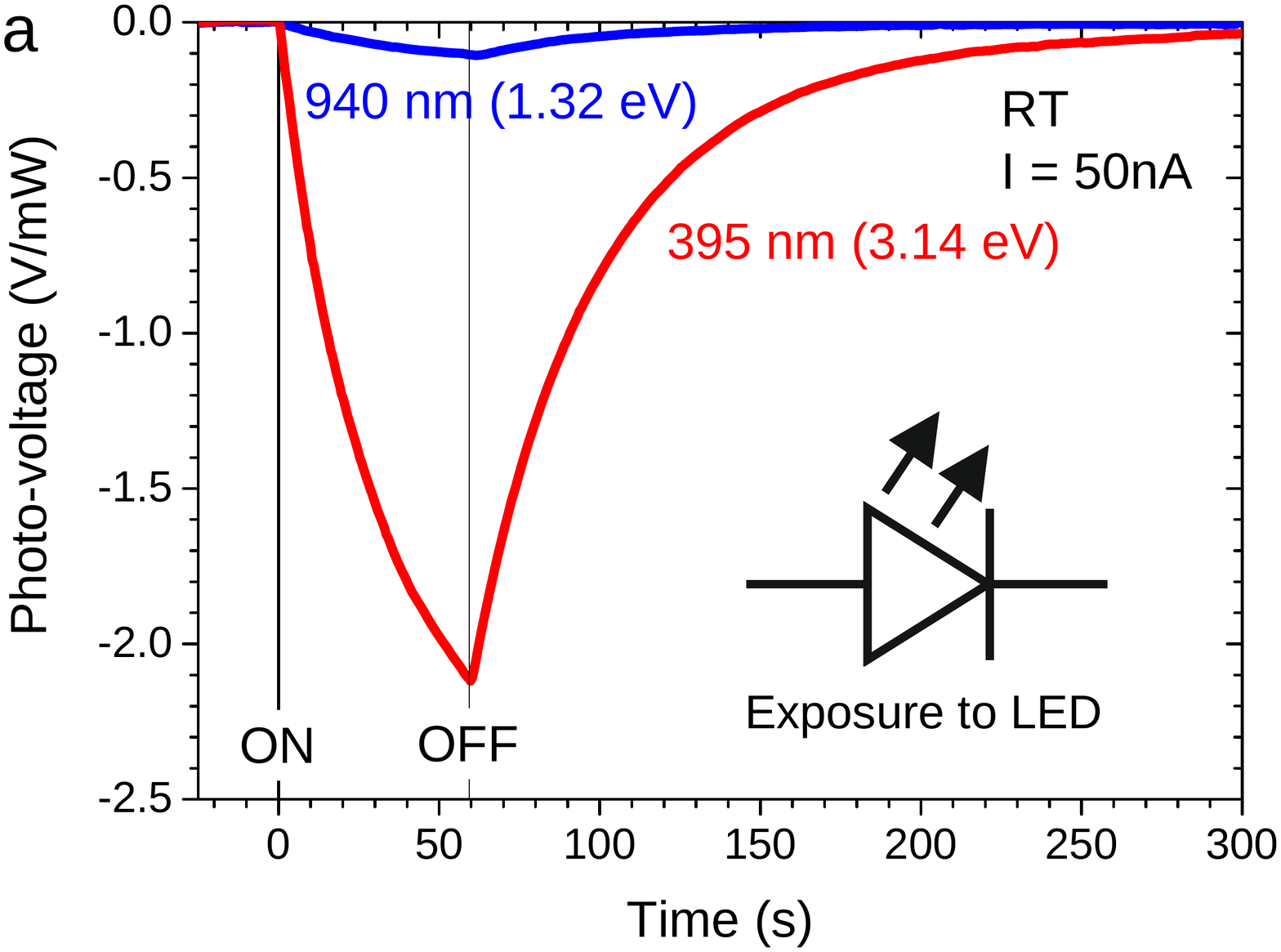}\includegraphics[width=80mm]{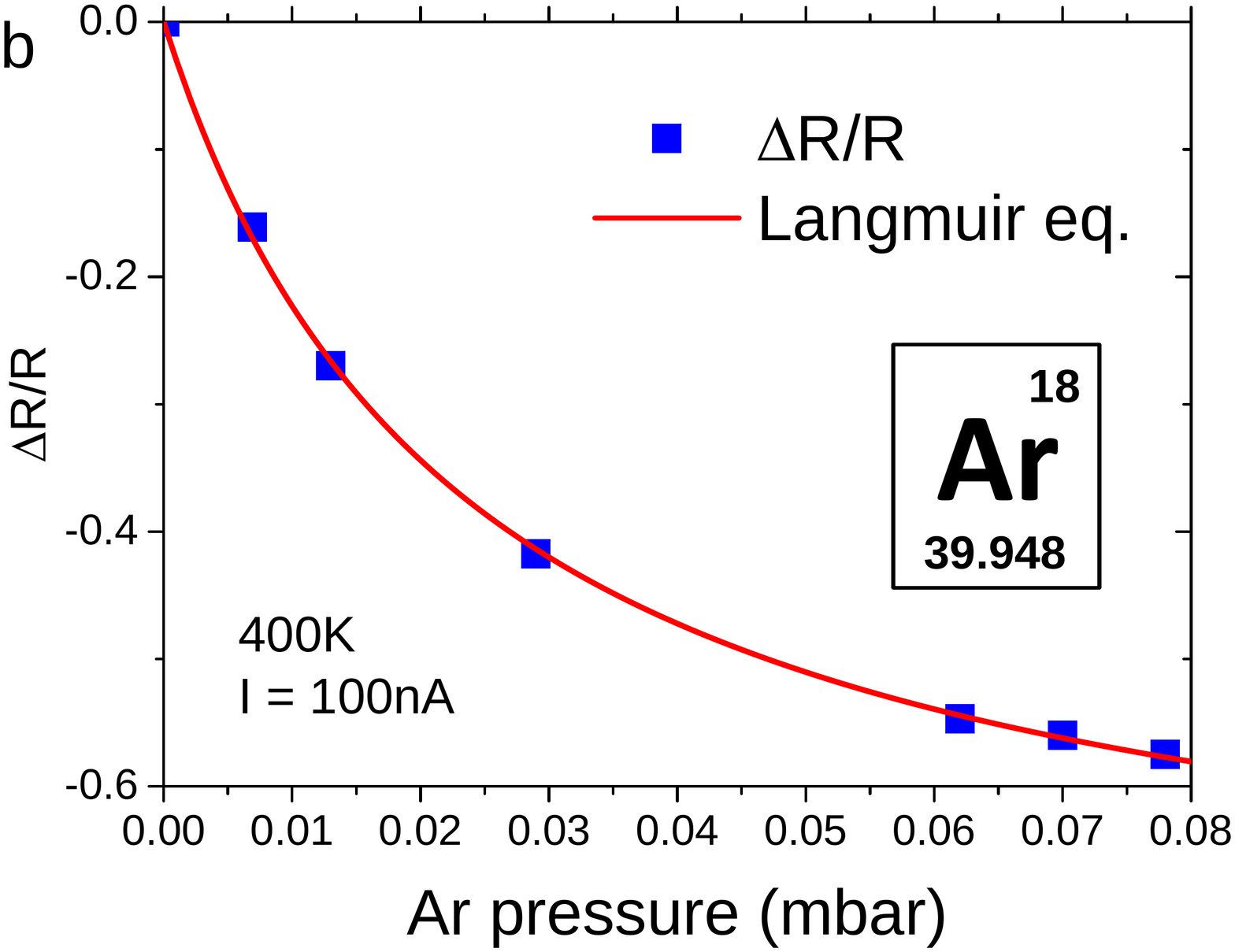}
\caption{{\bf a}. Normalized photo-voltage (V/mW) upon UV/IR illumination measured at 50~nA at room temperature.
{\bf b}. Changes in resistance vs argon pressure measured at 100~nA at 400~K.}
\label{Sensing}
\end{figure}

Our TCNQ--MOF with the enhanced electrical conduction is a good candidate as a resistive sensing element. Figure \ref{Sensing}a demonstrates the resistance at room temperature markedly reduced upon exposure to a UV light-emitting diode (LED) radiation with a central wavelength (photon energy) of 395~nm (3.14~eV), while only a subtle reduction (about twenty times smaller) is observed with a infra-red LED radiation with a central wavelength (photon energy) of 940~nm (1.32~eV). This is fully consistent with a optical absorption gap of $\sim$1.5~eV.

We further test our device for gas sorption. Figure \ref{Sensing}b shows the resistance of the TCNQ--MOF at 400~K at different pressures of argon exhibiting its drop as the pressure increases. The response is reversible so that the resistance goes back to the original value when the gas is evacuated. The fact, that the reduction occurs in the same way upon exposures to argon, nitrogen, oxygen and carbon dioxide regardless of their redox properties, means that it is pressure induced rather than carrier doping. Generally, dissipative heat loss due to an increased number of gaseous molecules in the vacuum chamber leads to enhanced resistances, opposite to what is observed here. The fact that the pressure (P) dependence follows Langmuir's surface coverage, $\theta = P/(P+P_0)$, where $P_0$ = 0.024~mbar, indicates that the device responds to the physical adsorption onto the TCNQ--MOF.

\subsection*{Conclusion}
Charge transfer between molecular components is central to metallic conduction and superconductivity in charge-transfer complexes and doped organic molecules \cite{HADDON91N,HEBARD91N,TANIGAKI91N}. We have shown that the guest-host charge transfer and orbital hybridization induce the electrical conduction of MOFs, based on which we have performed 'resistive' photo- and gas-sensing experiments. The implementation of doped MOFs as highly porous electrodes demonstrated in this study would pave the way towards advanced technology that exploits extraordinary physical and chemical properties of MOFs.

\section*{Acknowledgements}
We thank S. Loyer and A. Stangl for technical assistance. This work was supported by the Austrian Science Fund (FWF), projects P621333-N20 and P27769-N20.

\section*{Author contributions}
H.S. conceived, designed and conducted the work. H.S. prepared the samples. H.P. set up and H.P. and H.S. carried out the XRD measurements. B.C.B and J.C.M set up and carried out the TEM. H.S. and W.L. set up and carried out the transport experiments. T.P. set up and H.S. carried out the absorption, Raman and XPS.

\section*{Competing financial interests}
The authors declare no competing financial interests.

\clearpage

\clearpage

\begin{thebibliography}{10}

\bibitem{Yaghi03N}
O.~M. Yaghi, M.~O'Keeffe, N.~W. Ockwig, H.~K. Chae, M.~Eddaoudi, and J.~Kim.
\newblock Reticular synthesis and the design of new materials.
\newblock {\em Nature}, 423(6941):705--714, June 2003.

\bibitem{Kitagawa04ACE}
S.~Kitagawa, R.~Kitaura, and S.~Noro.
\newblock Functional porous coordination polymers.
\newblock {\em Angewandte Chemie-international Edition}, 43(18):2334--2375,
  2004.

\bibitem{Allendorf15JoPCL}
M.~D. Allendorf, M.~E. Foster, F.~Leonard, V.~Stavila, P.~L. Feng, F.~P. Doty,
  K.~Leong, E.~Y. Ma, S.~R. Johnston, and A.~A. Talin.
\newblock Guest-induced emergent properties in metal-organic frameworks.
\newblock {\em Journal of Physical Chemistry Letters}, 6(7):1182--1195, April
  2015.

\bibitem{Deria14CSR}
P.~Deria, J.~E. Mondloch, O.~Karagiaridi, W.~Bury, J.~T. Hupp, and O.~K. Farha.
\newblock Beyond post-synthesis modification: evolution of metal-organic
  frameworks via building block replacement.
\newblock {\em Chemical Society Reviews}, 43(16):5896--5912, August 2014.

\bibitem{Li12CR}
J.~R. Li, J.~Sculley, and H.~C. Zhou.
\newblock Metal-organic frameworks for separations.
\newblock {\em Chemical Reviews}, 112(2):869--932, February 2012.

\bibitem{Kreno12CR}
L.~E. Kreno, K.~Leong, O.~K. Farha, M.~Allendorf, R.~P. Van~Duyne, and J.~T.
  Hupp.
\newblock Metal-organic framework materials as chemical sensors.
\newblock {\em Chemical Reviews}, 112(2):1105--1125, February 2012.

\bibitem{Rosi03S}
N.~L. Rosi, J.~Eckert, M.~Eddaoudi, D.~T. Vodak, J.~Kim, M.~O'Keeffe, and O.~M.
  Yaghi.
\newblock Hydrogen storage in microporous metal-organic frameworks.
\newblock {\em Science}, 300(5622):1127--1129, May 2003.

\bibitem{Sumida12CR}
K.~Sumida, D.~L. Rogow, J.~A. Mason, T.~M. McDonald, E.~D. Bloch, Z.~R. Herm,
  T.~H. Bae, and J.~R. Long.
\newblock Carbon dioxide capture in metal-organic frameworks.
\newblock {\em Chemical Reviews}, 112(2):724--781, February 2012.

\bibitem{Furukawa10S}
H.~Furukawa, N.~Ko, Y.~B. Go, N.~Aratani, S.~B. Choi, E.~Choi, A.~O. Yazaydin,
  R.~Q. Snurr, M.~O'Keeffe, J.~Kim, and O.~M. Yaghi.
\newblock Ultrahigh porosity in metal-organic frameworks.
\newblock {\em Science}, 329(5990):424--428, July 2010.

\bibitem{Shimomura10NC}
S.~Shimomura, M.~Higuchi, R.~Matsuda, K.~Yoneda, Y.~Hijikata, Y.~Kubota,
  Y.~Mita, J.~Kim, M.~Takata, and S.~Kitagawa.
\newblock Selective sorption of oxygen and nitric oxide by an electron-donating
  flexible porous coordination polymer.
\newblock {\em Nature Chemistry}, 2(8):633--637, August 2010.

\bibitem{Shimomura07JotACS}
S.~Shimomura, S.~Horike, R.~Matsuda, and S.~Kitagawa.
\newblock Guest-specific function of a flexible undulating channel in a
  7,7,8,8-tetracyano-p-quinodimethane dimer-based porous coordination polymer.
\newblock {\em Journal of the American Chemical Society}, 129(36):10990--+,
  September 2007.

\bibitem{Chae04N}
H.~K. Chae, D.~Y. Siberio-Perez, J.~Kim, Y.~Go, M.~Eddaoudi, A.~J. Matzger,
  M.~O'Keeffe, and O.~M. Yaghi.
\newblock A route to high surface area, porosity and inclusion of large
  molecules in crystals.
\newblock {\em Nature}, 427(6974):523--527, February 2004.

\bibitem{Inokuma10NC}
Y.~Inokuma, T.~Arai, and M.~Fujita.
\newblock Networked molecular cages as crystalline sponges for fullerenes and
  other guests.
\newblock {\em Nature Chemistry}, 2(9):780--783, September 2010.

\bibitem{Stavila14CSR}
V.~Stavila, A.~A. Talin, and M.~D. Allendorf.
\newblock Mof-based electronic and optoelectronic devices.
\newblock {\em Chemical Society Reviews}, 43(16):5994--6010, August 2014.

\bibitem{Sheberla14JotACS}
D.~Sheberla, L.~Sun, M.~A. Blood-Forsythe, S.~Er, C.~R. Wade, C.~K. Brozek,
  A.~Aspuru-Guzik, and M.~Dinca.
\newblock High electrical conductivity in
  ni-3(2,3,6,7,10,11-hexaiminotriphenylene)(2), a semiconducting metal-organic
  graphene analogue.
\newblock {\em Journal of the American Chemical Society}, 136(25):8859--8862,
  June 2014.

\bibitem{Talin14S}
A.~A. Talin, A.~Centrone, A.~C. Ford, M.~E. Foster, V.~Stavila, P.~Haney, R.~A.
  Kinney, V.~Szalai, F.~El~Gabaly, H.~P. Yoon, F.~Leonard, and M.~D. Allendorf.
\newblock Tunable electrical conductivity in metal-organic framework thin-film
  devices.
\newblock {\em Science}, 343(6166):66--69, January 2014.

\bibitem{Rosi05JotACS}
N.~L. Rosi, J.~Kim, M.~Eddaoudi, B.~L. Chen, M.~O'Keeffe, and O.~M. Yaghi.
\newblock Rod packings and metal-organic frameworks constructed from rod-shaped
  secondary building units.
\newblock {\em Journal of the American Chemical Society}, 127(5):1504--1518,
  February 2005.

\bibitem{Dietzel05ACE}
P.~D.~C. Dietzel, Y.~Morita, R.~Blom, and H.~Fjellvag.
\newblock An in situ high-temperature single-crystal investigation of a
  dehydrated metal-organic framework compound and field-induced magnetization
  of one-dimensional metaloxygen chains.
\newblock {\em Angewandte Chemie-international Edition}, 44(39):6354--6358,
  2005.

\bibitem{Diaz-Garcia14CG&D}
M.~Diaz-Garcia, A.~Mayoral, I.~Diaz, and M.~Sanchez-Sanchez.
\newblock Nanoscaled m-mof-74 materials prepared at room temperature.
\newblock {\em Crystal Growth \& Design}, 14(5):2479--2487, May 2014.

\bibitem{Mayoral15C}
A.~Mayoral, M.~Sanchez-Sanchez, A.~Alfayate, J.~Perez-Pariente, and I.~Diaz.
\newblock Atomic observations of microporous materials highly unstable under
  the electron beam: The cases of ti-doped alpo4-5 and zn-mof-74.
\newblock {\em Chemcatchem}, 7(22):3719--3724, November 2015.

\bibitem{Gimeno-Fabra12CC}
M.~Gimeno-Fabra, A.~S. Munn, L.~A. Stevens, T.~C. Drage, D.~M. Grant, R.~J.
  Kashtiban, J.~Sloan, E.~Lester, and R.~I. Walton.
\newblock Instant mofs: continuous synthesis of metal-organic frameworks by
  rapid solvent mixing.
\newblock {\em Chemical Communications}, 48(86):10642--10644, 2012.

\bibitem{BOYD65JoCP}
R.~H. BOYD and W.~D. PHILLIPS.
\newblock Solution dimerization of tetracyanoquinodimethane ion radical.
\newblock {\em Journal of Chemical Physics}, 43(9):2927--\&, 1965.

\bibitem{HIROMA71BotCSoJ}
S.~HIROMA, H.~KURODA, and H.~AKAMATU.
\newblock Polarized absorption spectra of single crystals of ion radical salts
  .2. k(tcnq) and cs2(tcnq)3.
\newblock {\em Bulletin of the Chemical Society of Japan}, 44(1):9--+, 1971.

\bibitem{Oohashi73BotCSoJ}
Yukako Oohashi and Tadayoshi Sakata.
\newblock The reflection spectra of simple salts of the
  tetracyanoquinodimethane anion radical.
\newblock {\em Bulletin of the Chemical Society of Japan}, 46(11):3330--3335,
  1973.

\bibitem{TORRANCE79PRB}
J.~B. TORRANCE, B.~A. SCOTT, B.~WELBER, F.~B. KAUFMAN, and P.~E. SEIDEN.
\newblock Optical-properties of the radical cation tetrathiafulvalenium (ttf+)
  in its mixed-valence and mono-valence halide salts.
\newblock {\em Physical Review B}, 19(2):730--741, 1979.

\bibitem{Hashimoto16CPL}
S.~Hashimoto, A.~Yabushita, T.~Kobayashi, and I.~Iwakura.
\newblock Real-time measurements of ultrafast electronic dynamics in the
  disproportionation of [tcnq](2-)(2) using a visible sub-10 fs pulse laser.
\newblock {\em Chemical Physics Letters}, 650:47--51, April 2016.

\bibitem{KHATKALE79JoCP}
M.~S. KHATKALE and J.~P. DEVLIN.
\newblock Vibrational and electronic-spectra of the monoanion, dianion, and
  trianion salts of tcnq.
\newblock {\em Journal of Chemical Physics}, 70(4):1851--1859, 1979.

\bibitem{SUCHANSKI76JotACS}
M.~R. SUCHANSKI and R.~P. VANDUYNE.
\newblock Resonance raman spectroelectrochemistry .4. oxygen decay chemistry of
  tetracyanoquinodimethane dianion.
\newblock {\em Journal of the American Chemical Society}, 98(1):250--252, 1976.

\bibitem{JONKMAN72CPL}
H.~T. JONKMAN and KOMMANDE.J.
\newblock Uv spectra and their calculation of tcnq and its monovalent and
  divalent anion.
\newblock {\em Chemical Physics Letters}, 15(4):496--\&, 1972.

\bibitem{BIEBER74CP}
A.~BIEBER and J.~J. ANDRE.
\newblock Some electronic properties of tmpd, ca and tcnq molecules and their
  monoions and diions.
\newblock {\em Chemical Physics}, 5(2):166--182, 1974.

\bibitem{Grossel00CoM}
M.~C. Grossel, A.~J. Duke, D.~B. Hibbert, I.~K. Lewis, E.~A. Seddon, P.~N.
  Horton, and S.~C. Weston.
\newblock An investigation of the factors that influence the decomposition of
  7,7 ',8,8 '-tetracyanoquinodimethane (tcnq) and its salts to, and structural
  characterization of, the alpha,alpha-dicyano-p-toluoylcyanide anion.
\newblock {\em Chemistry of Materials}, 12(8):2319--2323, August 2000.

\bibitem{MILLER87JoPC}
J.~S. MILLER, J.~H. ZHANG, W.~M. REIFF, D.~A. DIXON, L.~D. PRESTON, A.~H. REIS,
  E.~GEBERT, M.~EXTINE, J.~TROUP, A.~J. EPSTEIN, and M.~D. WARD.
\newblock Characterization of the charge-transfer reaction between
  decamethylferrocene and 7,7,8,8-tetracyano-para-quinodimethane (1-1) - the
  fe-57 mossbauer-spectra and structures of the paramagnetic dimeric and the
  metamagnetic one-dimensional salts and the molecular and
  electronic-structures of [tcnq]n (n = 0, -1, -2)s.
\newblock {\em Journal of Physical Chemistry}, 91(16):4344--4360, July 1987.

\bibitem{HARRIS95VS}
M.~HARRIS, J.~J. HOAGLAND, U.~MAZUR, and K.~W. HIPPS.
\newblock Raman and infrared-spectra of metal-salts of
  alpha,alpha-dicyano-p-toluoylcyanide - nonresonant raman-scattering in
  tetracyano-p-quinodimethanide.
\newblock {\em Vibrational Spectroscopy}, 9(3):273--277, September 1995.

\bibitem{Fujisawa15PCCP}
J.~Fujisawa and M.~Hanaya.
\newblock Extremely strong organic-metal oxide electronic coupling caused by
  nucleophilic addition reaction.
\newblock {\em Physical Chemistry Chemical Physics}, 17(25):16285--16293, 2015.

\bibitem{Tseng10NC}
T.~C. Tseng, C.~Urban, Y.~Wang, R.~Otero, S.~L. Tait, M.~Alcami, D.~Ecija,
  M.~Trelka, J.~M. Gallego, N.~Lin, M.~Konuma, U.~Starke, A.~Nefedov,
  A.~Langner, C.~Woll, M.~A. Herranz, F.~Martin, N.~Martin, K.~Kern, and
  R.~Miranda.
\newblock Charge-transfer-induced structural rearrangements at both sides of
  organic/metal interfaces.
\newblock {\em Nature Chemistry}, 2(5):374--379, May 2010.

\bibitem{Precht15PCCP}
R.~Precht, R.~Hausbrand, and W.~Jaegermann.
\newblock Electronic structure and electrode properties of
  tetracyanoquinodimethane (tcnq): a surface science investigation of lithium
  intercalation into tcnq.
\newblock {\em Physical Chemistry Chemical Physics}, 17(9):6588--6596, 2015.

\bibitem{ROSE55PR}
A.~ROSE.
\newblock Space-charge-limited currents in solids.
\newblock {\em Physical Review}, 97(6):1538--1544, 1955.

\bibitem{HADDON91N}
R.~C. HADDON, A.~F. HEBARD, M.~J. ROSSEINSKY, D.~W. MURPHY, S.~J. DUCLOS, K.~B.
  LYONS, B.~MILLER, J.~M. ROSAMILIA, R.~M. FLEMING, A.~R. KORTAN, S.~H. GLARUM,
  A.~V. MAKHIJA, A.~J. MULLER, R.~H. EICK, S.~M. ZAHURAK, R.~TYCKO, G.~DABBAGH,
  and F.~A. THIEL.
\newblock Conducting films of c60 and c70 by alkali-metal doping.
\newblock {\em Nature}, 350(6316):320--322, March 1991.

\bibitem{HEBARD91N}
A.~F. HEBARD, M.~J. ROSSEINSKY, R.~C. HADDON, D.~W. MURPHY, S.~H. GLARUM,
  T.~T.~M. PALSTRA, A.~P. RAMIREZ, and A.~R. KORTAN.
\newblock Superconductivity at 18-k in potassium-doped c-60.
\newblock {\em Nature}, 350(6319):600--601, April 1991.

\bibitem{TANIGAKI91N}
K.~TANIGAKI, T.~W. EBBESEN, S.~SAITO, J.~MIZUKI, J.~S. TSAI, Y.~KUBO, and
  S.~KUROSHIMA.
\newblock Superconductivity at 33-k in csxrbyc60.
\newblock {\em Nature}, 352(6332):222--223, July 1991.

\end{thebibliography}

\end{document}